\documentclass{PoS}
\usepackage{epsf,epsfig,float,amssymb,latexsym,amsmath,amsthm,fancyhdr}
\usepackage{slashed}
\newcommand{\Deltaless}{\slashed{\Delta}}
\newcommand{\sigmaterm}{\sigma_{\pi N}}

\title{Low energy analysis of $\pi N$ scattering\\ and the pion-nucleon sigma term with \\ covariant baryon chiral perturbation theory}

\ShortTitle{$\pi N$ scattering and the pion-nucleon sigma term}

\author{\speaker{J. M. Alarc\'on}\\
    Institut f\"ur Kernphysik, Johannes Gutenberg Universit\"at, Mainz D-55099, Germany    \\
        E-mail: \email{alarcon@kph.uni-mainz.de}}

\author{J. Martin Camalich\\
        Department of Physics and Astronomy, University of Sussex, BN1 9QH, Brighton, UK\\
        E-mail: \email{J.Camalich@sussex.ac.uk}}
        
\author{J. A. Oller\\
       Departamento de F\'{\i}sica. Universidad de Murcia. E-30071, Murcia, Spain \\
        E-mail: \email{oller@um.es}}

\abstract{The pion-nucleon sigma term ($\sigmaterm$) is an observable of fundamental importance because embodies information about the internal scalar structure of the nucleon. Nowadays this quantity has triggered renewed interest because it is a key input for a reliable estimation of the dark matter-nucleon spin independent elastic scattering cross section. In this proceeding we present how this quantity can be reliably extracted by employing only experimental information with the use covariant baryon chiral perturbation theory. We also contrast our extraction with updated phenomenology related to $\sigmaterm$ and show how this phenomenology favours a relatively large value of $\sigmaterm$. Finally, we extract a value of $\sigmaterm=59(7)$~MeV from modern partial wave analyses data.}

\FullConference{Xth Quark Confinement and the Hadron Spectrum,\\
		October 8-12, 2012\\
		TUM Campus Garching, Munich, Germany}

\begin{document}

\section{Introduction}

Pion-nucleon scattering is a fundamental reaction that gives access to fundamental questions related to strong interactions. At higher energies, for example, allow us to study the baryonic spectrum of QCD and its properties. On the other hand, at low energies it is an excellent test for the chiral dynamics of QCD , able to provide a systematic framework to study isospin violation as well as valuable information about the internal structure of the nucleon. Regarding the latter, there is a strong demand from the dark matter community for an accurate value of the scalar coupling of the nucleon at zero momentum transfer, which is the definition of the pion-nucleon sigma term ($\sigmaterm$). This is so because there are important discrepancies between the values for $\sigmaterm$  
obtained by different partial wave analyses (PWAs). Namely, the classical partial wave analysis (PWA) of the Karlsruhe group gives a value of $\sigmaterm \approx 45(8)$~MeV \cite{KA85}, while the more modern PWA of the George Washington University group obtains $\sigmaterm \approx 64(7)$~MeV \cite{WI08}. This disagreement on the sigma-term value is the main hadronic uncertainty for the direct detection of dark matter \cite{hadronic-uncertainties}. 

Chiral perturbation theory (ChPT) is an excellent tool that can shed light on this issue because it provides a systematic and model independent way to study perturbatively the strong interactions of hadrons. From this perturbative construction one can also include formally, in a quantum field approach, the interaction with scalar sources \cite{gasser1}, from which the scalar form factor of the nucleon can be investigated. In fact, ChPT has been used several times to estimate the value of the pion-nucleon sigma term (see Refs.~\cite{fettes3, fettes_ep}), although the poor convergence of the chiral series prevented to make any conclusion about its value. 
As we show here, the lack of convergence of the previous analyses can be overcome if one works within a framework where one preserves the good analytical properties of a covariant calculation and includes the relevant degrees of freedom for the process considered. Namely, for $\pi N$ scattering, it is known that the $\Delta(1232)$ resonance plays a fundamental role because its closeness to the $\pi N$ threshold.
This makes that the behaviour of this resonance cannot be faithfully reproduced by a finite polynomial via the resonance saturation hypothesis \cite{ecker}. In our work we considered explicitly the contributions of the $\Delta(1232)$ following \cite{Pascalutsa:1998pw}. 

One added difficulty that one should overcome when working with covariant baryon chiral perturbation theory is the breaking of the perturbative expansion already addressed in Ref.~\cite{gasser2}. This problem is solved when working within the extended-on-mass-shell scheme (EOMS), which cancels the PCBT (which are analytical pieces) by renormalizing the low-energy counterterms (LECs) of the most general chiral Lagrangian \cite{eoms1, eoms2}.  

One of the main advantages of this scheme respect to the other covariant approach, infrared regularization (IR) \cite{becher}, is that EOMS preserves the good analytical properties of the relativistic scattering amplitudes. However, IR gives rise to unphysical cuts that limit the convergence of the chiral amplitude, as shown in \cite{becher,proceedings, nuestroEOMS}. For this reason, we decided to work within the EOMS to preserve the good analytical properties of our calculated amplitudes as well as the standard power counting of ChPT. The results presented here are part of the detailed analysis of the $\pi N$ scattering phenomenology of Ref.~\cite{nuestroEOMS}, where the $\pi N$ scattering amplitude is calculated up to $\mathcal{O}(p^3)$ including the $\Delta(1232)$. 


\newpage

\section{Fits}
\vspace{-0.6cm}
In order to determine the LECs, which numerical values are not fixed by chiral symmetry, we decided to fit our chiral amplitude to the phase shifts provided by three different PWAs. These are the PWAs of the Karlsruhe group (KA85) \cite{KA85}, the George Washington University group (WI08) \cite{WI08} and the Matsinos' group (EM06) \cite{EM06}.   

\begin{figure}[H]
\centerline{\epsfig{file=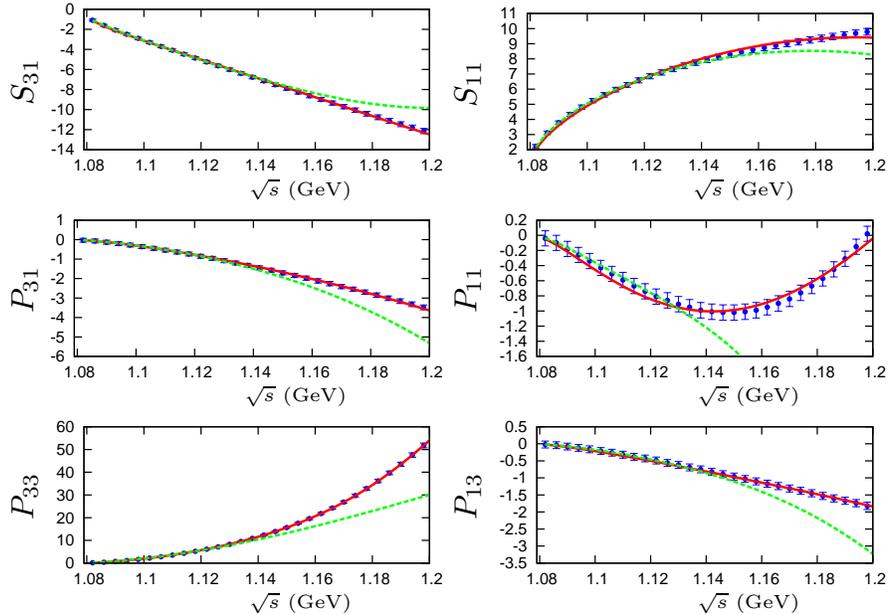,width=.77\textwidth,angle=0}}
\caption[pilf]{\protect \small Fits to WI08~\cite{WI08} with (solid red line) and without (green dashed line) the inclusion of the $\Delta(1232)$ as an explicit degree of freedom. For \label{fitsWI08}}
\end{figure}

In Fig. \ref{fitsWI08} we observe that in the $\Deltaless$-ChPT case (green dashed line) we are able to fit well the WI08 phase shifts up to energies of $\sqrt{s}=1.14$-$1.16$~GeV, depending of which partial wave one consider. This situation is considerably improved once the $\Delta(1232)$ is included explicitly in our formalism  $\Delta$-ChPT (red solid line), in which case we are able to describe the data very well up to energies of $\sqrt{s}=1.20$~GeV. Although not shown here, this good agreement is also observed for KA85 and EM06 (see Ref.~\cite{nuestroEOMS}).


The values of the LECs extracted from these fits are shown in Table~\ref{LECs}. The results of the $\Deltaless$-ChPT are given in columns 5-7, while the results of the $\Delta$-ChPT fits are shown in columns 2-4. In the latter case we have one extra LEC compared to the $\Deltaless$-ChPT, the $N\Delta$ axial coupling ($h_A$). This LEC, on the other hand, can be related to the $\Delta(1232)$ Breit-Wigner width, $\Gamma_\Delta$. So, taking as input the value reported by the PDG $\Gamma_\Delta=118(2)$~MeV \cite{PDG}, one deduces a value of $h_A=2.90(2)$. However, we decided to leave this LEC as a free parameter to check the reliability of the different PWAs.
Once we have determined completely the chiral amplitude, a key point is to predict other interesting phenomenology. Here we present  the results regarding the convergence of the chiral amplitude in the subthreshold region, Sec.~\ref{Sec:subthreshold} and  the extraction of $\sigmaterm$, Sec.~\ref{piNsigmaterm}


\begin{table}[H]\footnotesize
 \begin{center}
\begin{tabular}{|r|r|r|r|r|r|r|}
\hline
\small{LEC}       &   KA85	       &   WI08	   & EM06	                        & KA85  	      &    WI08 	& EM06  				   \\
                  &	$\Delta$-ChPT    &	$\Delta$-ChPT	   &	$\Delta$-ChPT	&  $\slashed{\Delta}$-ChPT & $\slashed{\Delta}$-ChPT  & $\slashed{\Delta}$-ChPT  	   \\
\hline		   	   																	      
$c_1$             &  -0.80(6)   & -1.004(30)  &  -1.000(8)            &  $-1.26(14)$   &  $-1.50(7)$  & $-1.47(2) $			     \\   
$c_2$             &  1.12(13)   &  1.010(40)  &  0.575(25)             &  $ 4.08(19) $  &  $3.74(26)$   & $3.63(2) $			     \\
$c_3$             &  -2.96(15)  & -3.040(20)  &  -2.515(35)            &  $-6.74(38)$  &  $-6.63(31)$  & $-6.42(1) $			      \\
$c_4$             &  2.00(7)    & 2.029(10)   &   1.776(20)             &  $3.74(16)$    &  $3.68(14)$   & $3.56(1)$			        \\
\hline		   					   													      
$d_1+d_2$         &  -0.15(21)  &   0.15(20)  &    -0.34(5)         &  $3.3(7)$    &  $3.7(6)$   &  $ 3.64(8)$			      \\
$d_3$             &  -0.21(26)  &  -0.23(27)   &   0.276(43)         & $-2.7(6)$   &  $-2.6(6)$  & $-2.21(8) $			      \\
$d_5$             &   0.82(14) &   0.47(7)  & 	    0.2028(33)     & $0.50(35) $    &  $-0.07(16)$  & $ -0.56(4)$			      \\
$d_{14}-d_{15}$   &   -0.11(44)  &  -0.5(5)  &     0.35(9)       & $-6.1(1.2)$    &  $-6.8(1.1)$  & $ -6.49(2)$			        \\
 $d_{18}$         &   -1.53(27)  &  -0.2(8)  &     -0.53(12)    & $-3.0(1.6)$    &  $-0.50(1.8)$  & $ -1.07(22)$			            \\ 
\hline  
$h_A$             &  $3.02(4)$   &  $2.87(4)$  & $2.99(2)$             &	 -- 	&     --	      & 	-- 		               \\
\hline
$\chi^2_{\rm d.o.f.}$  &       $0.77$       &     $0.24$          & $0.11$    &	 $0.38$ 	&     $0.23$	      & 	$25.08$ \\ 
\hline 
\end{tabular}
\caption[pilf]{\protect \small This table gathers the values of the LECs extracted in the different fits, with and without the $\Delta(1232)$. The $\mathcal{O}(p^2)$ and $\mathcal{O}(p^3)$ LECs (the $c_i$ and $d_i$) are shown in units of $GeV^{-1}$ and $GeV^{-2}$ respectively. \label{LECs}}
\end{center}
\end{table}

\section{Subthreshold region}
\label{Sec:subthreshold}

The subthreshold region is of special interest because contains points that are related to important low energy theorems. 
One of the most important and known of these theorems in $\pi N$ scattering is the Cheng-Dashen theorem \cite{cheng-dashen-theorem}, that relates the value of the Born-subtracted isoscalar scattering amplitude, $\bar{D}^+$, at the Cheng-Dashen point, to the scalar form factor of the nucleon, $\sigma(t)$, at $t=2 M_\pi^2$. 
This theorem is used by several PWAs to extract the value of $\sigmaterm$, since the difference $\sigma(2 M_\pi^2)-\sigmaterm\equiv \Delta_\sigma$ can be derived in different ways, see for example Refs.~\cite{gasser2,beche2,formfactorsigmaterm}.   
A difficulty one had in the subthreshold region is that the ChPT analyses of $\pi N$ scattering could not extract, from physical information, the value of the subthreshold quantities that the PWAs are able to obtain by means of dispersive methods. 
Therefore a good way to explore the convergence of the chiral series in this region is to compare our results there with the one obtained by their corresponding PWAs. The subthreshold quantities that we studied are the subthreshold coefficients $\bar{d}_{00}^+$ and $\bar{d}_{01}^+$, defined by the power expansion $\bar{D}^+(\nu=0,t)=\bar{d}_{00}^+  + \bar{d}_{01}^+ t + \dots$ and the so called $\Sigma$-term, $\Sigma \equiv f_\pi^2 \bar{D}^+(\nu=0,t=2M_\pi^2)$. We chose these quantities because they are closely connected to the value of $\sigmaterm$ \cite{nuestroEOMS}.

\begin{table}[h!]\footnotesize
 \begin{center}
\begin{tabular}{|c|c|c|c||c|c|c||c|c|}
\hline
                                     &   KA85 \cite{nuestroEOMS} &  WI08  \cite{nuestroEOMS}     & EM06  \cite{nuestroEOMS}      & KA85  \cite{nuestroEOMS} & WI08  \cite{nuestroEOMS} & EM06  \cite{nuestroEOMS}	       &  KA85  & WI08    \\
                                     &  $\Deltaless$-ChPT        &    $\Deltaless$-ChPT            & $\Deltaless$-ChPT              &  $\Delta$-ChPT      &    $\Delta$-ChPT       &    $\Delta$-ChPT              & \cite{KA85} &\cite{WI08} \\ 
\hline
 $d_{00}^+$ ($M_\pi^{-1}$)            &  $-2.02(42)$ & $-1.65(28)$ & $-1.56(5)$ & $-1.48(15)$ & $-1.20(13)$ &  $-0.97(2)$    &   $-1.46$    & $-1.30$                 \\
 $d_{01}^+$ ($M_\pi^{-3}$)            &  $1.73(19)$  & $1.70(18)$  & $1.64(4)$ & $1.21(10)$ & $1.20(9)$ & $1.08(2)$         &    $1.14$    &  $1.19$           \\ 
 $\Sigma$ (MeV)                     &   $84(10)\footnotemark[1]$  & $103(5)\footnotemark[1]$    & $103(2)\footnotemark[1]$  &  $45(7)\footnotemark[1]$& $64(6)\footnotemark[1]$ & $64(1)\footnotemark[1]$              &   $64(8)$    &  $79(7)$              \\ 
\hline		   	   
\end{tabular}
\caption[pilf]{\protect \small Values of the subthreshold quantities under consideration, extracted from direct extrapolation of the chiral amplitude into the subthreshold region. They are compared with the values reported by the PWAs (last two columns). \label{subthreshold}}
\end{center}
\end{table}

In Table \ref{subthreshold} we show the result of the extractions of $\bar{d}_{00}^+$, $\bar{d}_{01}^+$ and $\Sigma$.
We show there that, in the $\Delta$-less case, the general trend is to overestimate, in modulus, the value of these subthreshold quantities compared with those calculated by the corresponding PWAs. However, including the $\Delta$ as an explicit degree of freedom, we are able to extract values of the subthreshold coefficients and the $\Sigma$-term that are in agreement with the PWAs. This suggest that the $\Delta(1232)$ is a key ingredient for the convergence of the chiral series also in the subthreshold region. Once this resonance is included as an explicit degree of freedom we could, for the first time in the literature, connect the information that lies in the physical region, encoded in the LECs through our fits, with the amplitudes in the subthreshold region. This information was, up to now, only accessible by dispersive methods. However, the $\Delta(1232)$ alone is not enough to achieve the best convergence, and is also necessary to work in a  formalism where we preserve the good analytical properties of a covariant calculation to extract all the potential of baryon ChPT (see Ref.~\cite{nuestroEOMS} for more details). 

\footnotetext[1]{These values of $\Sigma$ are extracted from an $\mathcal{O}(p^3)$ ChPT calculation, which is known that underestimates the value of this quantity in approximately $10$~MeV \cite{gasser2, formfactorsigmaterm}. As discussed in Ref.[14] what matters for determining $\sigma_{\pi N}$ is $\Sigma_d=f_\pi^2(\bar{d}^+_{00}+2 M_\pi^2\bar{d}^+_{01})$.} 

\section{The pion-nucleon sigma term}
\label{piNsigmaterm} 

The pion-nucleon sigma term is an important quantity that contains information about the internal scalar structure of the nucleon.
It is related to the origin of the mass of ordinary matter and is important for the investigation of QCD phase diagram. In addition, nowadays it is widely used for estimations of the dark matter-nucleon spin independent elastic scattering cross section, which are used for direct detection of dark matter. 

The PWAs can extract the value of $\sigmaterm$ by extrapolating the $\bar{D}^+$ to the Cheng-Dashen point ($\nu=0, t=2 M_\pi^2$).\footnote[2]{The variable $\nu$ is defined in terms of the Mandelstam variables $s$ and $u$ as $\nu=\frac{s-u}{4 m_N}$.} However, this is a delicate extrapolation into the subthreshold region with no direct physical information to compare with. In this regard, the advantage of ChPT over dispersive methods is that chiral symmetry allow us to relate $\sigmaterm$ to the LEC $c_1$ (in an $\mathcal{O}(p^3)$ calculation) which, on the other hand, can be determined from experimental information. That means that once the value of $c_1$ is reliably fixed employing experimental information, and the convergence of the chiral expansion for $\sigmaterm$ is proven, one may give a reliable extraction of $\sigmaterm$.

The explicit relation between $\sigmaterm$ and $c_1$ can be obtained directly form $\sigma(t=0)$ or by applying the Hellmann-Feynman theorem on the chiral expansion of the nucleon mass. By both means one obtains, in an $\mathcal{O}(p^3)$ covariant calculation, the result\footnote[3]{In Eq.~\eqref{Eqsigmaterm}, $c_1$ corresponds to the EOMS renormalized LEC.} \cite{nuestrosigmaterm}:

\begin{align}\label{Eqsigmaterm}
\sigmaterm =  -4c_1 M_\pi^2-\frac{3 g_A^2 M_\pi^3}{16 \pi^2 f_\pi^2 m_N}  \left( \frac{3m_N^2-M_\pi^2}{\sqrt{4 m_N^2-M_\pi^2}}\arccos\frac{M_\pi}{2 m_N}+M_\pi \log \frac{M_\pi}{m_N} \right)
\end{align}

The point here is that our good convergence of the chiral series above and below threshold, once the $\Delta(1232)$ is included, give us confidence about the reliability of the LECs extracted with $\Delta$-ChPT from PWA phase shifts. On the other hand, the chiral expansion for $\sigmaterm$ exhibits a good convergence up to the order that we calculate, as was already shown in Ref.~\cite{nuestrosigmaterm}. Using the values of $c_1$ extracted with $\Delta$-ChPT, one obtains the values for $\sigmaterm$ shown in Table~\ref{sigmatermtable}. 

\begin{table}[H]\small
 \begin{center}
\begin{tabular}{|c|c|c|c||c|c|c|}
\hline
                        & KA85 \cite{nuestroEOMS} & WI08 \cite{nuestroEOMS} & EM06 \cite{nuestroEOMS}& KA85  & WI08 & EM06   \\
                        &   $\Delta$-ChPT   &  $\Delta$-ChPT    & $\Delta$-ChPT     & \cite{KA85} & \cite{WI08}  & \cite{EM06}            \\
\hline
 $c_1$ (GeV$^{-1}$)     &   $-0.80(6)$  & $-1.00(4)$      & $-1.00(1)$         &  --   &  --   &    --    \\
\hline
  $\sigmaterm$ (MeV)    & $43(5)$ & $59(4)$& $59(2)$ & $45(8)$ & $64(7)$ & $56(9)$                   \\
\hline		   	   
\end{tabular}
\caption[pilf]{\protect \small Values of $\sigmaterm$ extracted from the fits to the different PWAs. \label{sigmatermtable}}
\end{center}
\end{table}

As one can see in Table~\ref{sigmatermtable}, the $\Delta$-ChPT extraction is in good agreement with the one of the PWAs. This suggests that the discrepancy between KA85 and WI08 regarding the value of $\sigmaterm$ is not due the methodology used by these PWAs. Moreover, we observe that our extractions employing the modern PWAs WI08 and EM06 agree remarkably well, despite both follow a very different systematics. However, what WI08 and EM06 have clearly in common is that both employ modern and high quality data. This suggests that the modern data point to a relatively large value of $\sigmaterm$.

\begin{figure}[h]
\begin{center}
\epsfig{file=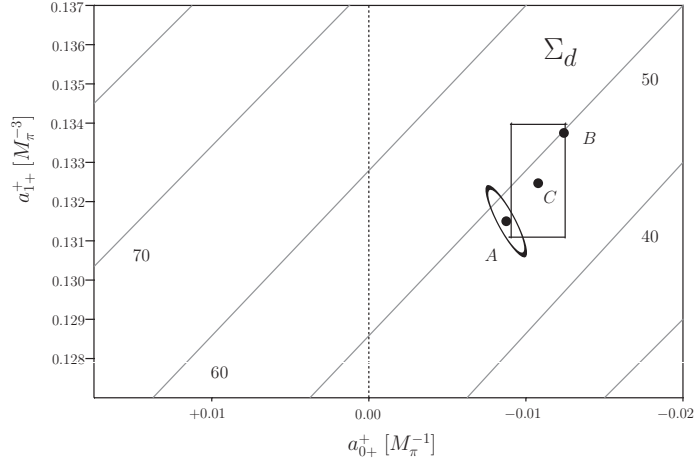,width=0.60\textwidth,angle=0}
\caption[pilf]{\protect \small Dependence of $\Sigma_d$ (diagonal lines) with with $a_{0+}^+$ and $a_{1+}^+$. 
The point A correspond to the values of  $a_{0+}^+$ and $a_{1+}^+$ deduced from the data of Bertin {\itshape et. al.} \cite{bertinetal} with one standard deviation (ellipse). The point C and the square correspond to the central values and errors for  $a_{0+}^+$ and $a_{1+}^+$ reported by KA85.
\label{sigma_dfigure}}
\end{center}
\end{figure}

In order to discriminate which extractions are more reliable we decided to calculate different quantities that can be compared with independent determinations. Two of those quantities are $h_A$ and the Goldberger-Treiman deviation, $\Delta_{GT}$. For $h_A$ we showed that only using the WI08 solution to fix the LECs, we extract a value for this coupling that is related to a $\Gamma_\Delta$ compatible with the value reported by the PDG. In the case of $\Delta_{GT}$, our extractions are compatible with independent determinations from $NN$ scattering and pionc-atoms only if we employ the modern PWAs of WI08 and EM06. 
Nevertheless, the most important quantity to compare with is maybe the scalar-isoscalar scattering length ($a_{0+}^+$), since this quantity is closely related to the value of $\sigmaterm$. In order to illustrate this, we show in Fig.~\ref{sigma_dfigure} the dependence of $\Sigma_d$ ($\Sigma_d - \sigmaterm \approx 3$~MeV)  as a function of $a_{0+}^+$ and $a_{1+}^+$. This relation was already shown in Ref.~\cite{sigmatermupdate}, from where Gasser, Leutwyler and Sainio deduced a value for $\sigmaterm$ of $45$~MeV. In fact, the value of $a_{0+}^+$ was key in their extraction (see Fig.~1 of that reference). There is clearly shown that the value of $\sigmaterm$ that they report (point A) relies on a negative value of $a_{0+}^+$, accepted by the time in which this paper was published. However, according to modern determinations for this scattering length extracted from pionic-atom \cite{baruNPA}, the value of $a_{0+}^+$ is now well consistent with zero. This affects directly to the value of $\sigmaterm$, which is pushed to larger values. This means that using {\it exactly the same argument} that Gasser, Leutwyler and Sainio employed to determine a value of $\sigmaterm\approx 45$~MeV would lead to a larger value of the sigma term in light of this updated determination. 

We also used the value of $a_{0+}^+$ extracted from $\pi$-atoms data to check the reliability of the PWAs employed in our extractions. As we show in Table~\ref{a0++Table}, we observe, again, that only using modern PWAs our extractions are more compatible with independent determinations. 

From these checks with independent and updated phenomenology, we conclude that the most reliable PWAs to use as input to extract $\sigmaterm$ are the modern PWAs of the George Washington University and Matsinos' groups. From these modern PWAs  we extract a value of $\sigmaterm=59(7)$~MeV, where the 7~MeV of error takes into account the uncertainty in the LEC $c_1$ and the first higher order correction to $\sigmaterm$ not taken into account in our $\mathcal{O}(p^3)$ calculation.

This value, unlike to the old value of 45~MeV, is consistent with the updated phenomenology related to the sigma term (see Ref,~\cite{nuestroEOMS, nuestrosigmaterm, nuestrosigmas}). This consistency with updated phenomenology is important to claim reliably for a certain value of $\sigmaterm$ and, unfortunately, is not satisfied by other experimental based determinations \cite{Stahov&Clement}. 
The more recent calculation of Ref.~\cite{EOMSchinos}, lacks also from consistency with independent determinations\footnote[5]{Namely, the $h_A$ extracted in this analysis is not compatible with the result deduced from the PDG. The threshold parameters, on the other hand, are not even considered in this work.} and extract results at odds with the PWA that they employ. 

\vspace{0.55cm}

\vspace{-0.5cm}
\begin{table}[H]
 \begin{center}
\begin{tabular}{|c|c|c|c|c|}
\hline
                                                                                       &  KA85 $\Delta$-ChPT \cite{nuestroEOMS}&   WI08 $\Delta$-ChPT \cite{nuestroEOMS}  &  EM06 $\Delta$-ChPT \cite{nuestroEOMS}  & $\pi$-atoms \cite{baruNPA}\\
\hline
 $a_{0+}^+$  ($10^{-3} M_\pi^{-1}$)   &  $-11(10)$                      & $-1.2(3.3)$               &           $2.3(2.0)$    &              $-1.0(9)$ \\
\hline
\end{tabular}
\caption[pilf]{\protect \small Values of $a_{0+}^+$ extracted from the different PWAs (columns 2-4) compared with the independent determination of Ref.~\cite{baruNPA}. The value shown in column 4 is obtained from Ref.~\cite{baruNPA} taking only $\pi^+ p$ and $\pi^- p$ scattering data.\label{a0++Table}}
\end{center}
\end{table}

\vspace{-1cm}

\section{Summary and Conclusions}

In this contribution we highlighted some of the results obtained in Ref.~\cite{nuestroEOMS}. There, we calculated the $\pi N$ scattering amplitude up to $\mathcal{O}(p^3)$ in the EOMS covariant scheme including the $\Delta(1232)$-resonance as a dynamical degree of freedom. We show how with this inclusion one achieves the best convergence in the physical and the subthreshold regions. Thanks to this improvement we were able to connect, for first time in the literature, the information extracted in the physical region with the one that lies in the subthreshold region.
This good convergence of the the chiral amplitude in both regions allowed us to extract reliably the value of the pion-nucleon sigma term form experimental information (PWAs), since the value of $\sigmaterm$ converges well in our covariant $\mathcal{O}(p^3)$ calculation including the $\Delta(1232)$ \cite{nuestroEOMS,nuestrosigmaterm,sigmatermproceedingJorge}. Employing the modern PWAs we extracted a value of $\sigmaterm=59(7)$~MeV, and compared this value with related phenomenology. We show how this relatively large value of $\sigmaterm$ is consistent with the updated phenomenology, while the old value of 45~MeV is only constent with a negative value of $a_{0+}^+$, which is at odds with modern $\pi$-atoms results. This consistency with updated phenomenology gives a strong support to the relatively large values of $\sigmaterm$, that has been recently shown to be also consistent with a negligible strange quark contribution to the nucleon \cite{nuestrosigmas}. These updated determinations of the sigma terms (Ref.~\cite{nuestrosigmaterm, nuestrosigmas}) from effective field theory can be very useful to the dark matter community.


\begin{thebibliography}{99}
\bibitem{KA85} R. Koch, Nucl. Phys. A {\bf 448} (1986) 707; R.~Koch and E.~Pietarinen, Nucl. Phys. A {\bf 336}  331 (1980).
   
   \bibitem{WI08} Computer code SAID, online program at http://gwdac.phys.gwu.edu/~, solution WI08. 
   R.~L.~Workman, {\it et al.}.
  Phys.\ Rev.\ C {\bf 86}, 035202 (2012).
   
   \bibitem{hadronic-uncertainties} 
   A.~Bottino, F.~Donato, N.~Fornengo and S.~Scopel,
  Astropart.\ Phys.\  {\bf 13}, 215 (2000);
  Astropart.\ Phys.\  {\bf 18}, 205 (2002).
  J.~R.~Ellis, K.~A.~Olive and C.~Savage,
  Phys.\ Rev.\ D {\bf 77}, 065026 (2008).
  
  \bibitem{gasser1} J.~Gasser and H.~Leutwyler,
  Annals Phys.\  {\bf 158} 142 (1984).

\bibitem{fettes3}N.~Fettes, U.-G.~Mei\ss ner and S.~Steininger,
  Nucl.\ Phys.\  A {\bf 640} 199 (1998).

\bibitem{fettes_ep} N.~Fettes and U.-G.~Mei\ss ner,
  Nucl.\ Phys.\  A {\bf 679} 629 (2001).
  
  \bibitem{ecker} 
  G.~Ecker, J.~Gasser, A.~Pich and E.~de Rafael,
  Nucl.\ Phys.\ B\ {\bf 321}, 311 (1989).

  \bibitem{Pascalutsa:1998pw}
  V.~Pascalutsa,
  Phys.\ Rev.\  D {\bf 58}, 096002 (1998);
  V.~Pascalutsa and R.~Timmermans,
  Phys.\ Rev.\  C {\bf 60}, 042201 (1999);  V.~Pascalutsa,
  Phys.\ Lett.\  B {\bf 503}, 85 (2001).

\bibitem{gasser2}
  J.~Gasser, M.~E.~Sainio and A.~Svarc,
  Nucl.\ Phys.\  B {\bf 307}  779 (1988).


\bibitem{eoms1}J.~Gegelia and G.~Japaridze,
  Phys.\ Rev.\  D {\bf 60}, 114038 (1999) .
\bibitem{eoms2} T.~Fuchs, J.~Gegelia, G.~Japaridze and S.~Scherer,
  Phys.\ Rev.\  D {\bf 68}, 056005 (2003).


\bibitem{becher}  T.~Becher and H.~Leutwyler,
  Eur.\ Phys.\ J.\  C {\bf 9} 643 (1999).
  
\bibitem{proceedings}  
  J.~M.~Alarcon, J.~M.~Camalich, J.~A.~Oller and L.~Alvarez-Ruso,
  Phys.\ Rev.\  C {\bf 83} 055205  (2011).
 ; arXiv:1107.3989 [nucl-th].
 ; AIP Conf.\ Proc.\  {\bf 1432}, 331 (2012).
; J.~M.~Alarcon, J.~Martin Camalich and J.~A.~Oller,
  Prog.\ Part.\ Nucl.\ Phys.\  {\bf 67}, 375 (2012).

\bibitem{nuestroEOMS} 
  J.~M.~Alarcon, J.~M.~Camalich and J.~A.~Oller,
  arXiv:1210.4450 [hep-ph].

\bibitem{beche2} T.~Becher and H.~Leutwyler,
  JHEP {\bf 0106} (2001) 017.
  
 
  
\bibitem{EM06}
  E.~Matsinos, W.~S.~Woolcock, G.~C.~Oades, G.~Rasche, A.~Gashi,
  Nucl.\ Phys.\  A {\bf 778 } 95 (2006).

  
\bibitem{PDG}
K.~Nakamura {\it et al.} [Particle Data Group Collaboration],
  J.\ Phys.\ G G\ {\bf 37} 075021 (2010).
  
    
  \bibitem{cheng-dashen-theorem}
  T.~P.~Cheng, R.~F.~Dashen,
  Phys.\ Rev.\ Lett.\  {\bf 26 }  594 (1971) .
  
  \bibitem{formfactorsigmaterm}
  J.~Gasser, H.~Leutwyler, M.~E.~Sainio,
  Phys.\ Lett.\ B {\bf 253}, 260 (1991).

  \bibitem{nuestrosigmaterm}
  J.~M.~Alarcon, J.~Martin Camalich, J.~A.~Oller,
  Phys.\ Rev.\ D {\bf 85}, 051503(R) (2012).

\bibitem{sigmatermupdate}
  J.~Gasser, H.~Leutwyler, M.~E.~Sainio,
  Phys.\ Lett.\ B {\bf 253}, 252 (1991).

 \bibitem{bertinetal}
P.~Y.~Bertin {\it et al.},
  Nucl.\ Phys.\ B {\bf 106}, 341 (1976).

  \bibitem{baruNPA} 
  V.~Baru, {\it et. al.},
  Nucl.\ Phys.\ A {\bf 872}, 69 (2011).

  \bibitem{nuestrosigmas}
  J.~M.~Alarcon, L. S. Geng, J.~Martin Camalich, J.~A.~Oller. arXiv:1209.2870 [hep-ph].

\bibitem{Stahov&Clement} 
  J.~Stahov, H.~Clement and G.~J.~Wagner,
  arXiv:1211.1148 [nucl-th].

\bibitem{EOMSchinos} 
  Y.~-H.~Chen, D.~-L.~Yao and H.~Q.~Zheng,
  arXiv:1212.1893 [hep-ph].

\bibitem{sigmatermproceedingJorge} 
  J.~Martin Camalich, J.~M.~Alarcon and J.~A.~Oller,
  Prog.\ Part.\ Nucl.\ Phys.\  {\bf 67}, 327 (2012).

\end{thebibliography}
\end{document}